# Density dependence of elastic properties of graphynes

Guilherme B. Kanegae and Alexandre F. Fonseca

*Applied Physics Department, "Gleb Wataghin" Institute of Physics, University of Campinas - UNICAMP, Campinas, São Paulo, CEP 13083-859, Brazil.*

## ABSTRACT

*Graphyne is a two-dimensional carbon allotrope of graphene. Its structure is composed of aromatic rings and/or carbon-carbon bonds connected by one or more acetylene chains. As some graphynes present the most of the excellent properties of graphene and non-null bandgap, they have been extensively studied. Recently, Kanegae and Fonseca reported calculations of four elastic properties of 70 graphynes, ten members of the seven families of graphynes [Carbon Trends 7, 100152 (2022)]. They showed that the acetylene chain length dependence of these properties can be simply modelled by a serial association of springs. Here, based on those results, we present the density dependence of these properties and show that the elastic moduli, $E$, of graphyne are less dependent on density, $\rho$, than porous cellular materials with an exponent of $E \sim \rho^n$, smaller than 2. We discuss the results in terms of the shape of the pores of the graphyne structures.*

## INTRODUCTION

Like graphene, graphyne (GY) is a bidimensional one-atom-thick carbon nanostructure. However, different from graphene, the structure of GY is composed of aromatic rings or carbon-carbon bonds connected by acetylene chains of length $n$, (–[–C≡C–]$_n$–) [1-3]. Like graphene, some GYs possess good electronic, mechanical and thermal properties [4-8], however, different from graphene, some GYs possess non-null band-gap [1-3,9]. Although the synthesis of GYs with $n = 2$ [10-12] and $n = 4$ [13,14] have been already reported in the literature, only recently the synthesis of GYs with $n = 1$ was achieved [15-18], boosting the scientific interest on these carbon nanostructures.

The combination of good physical properties and porosity make GYs to be promising structures to storage hydrogen [19], to filter water [20], to act as sensors [21], supercapacitors [22], antibacterial agents [23], amongst others [2,3].

Recently, we have performed several molecular dynamics (MD) simulations of the elastic properties of a total of seventy GY structures [24], ten members of each of the original seven types or families of GYs as proposed by Baughman, Eckhardt and Kertesz [1] in the eighties. The elastic properties obtained for each GY structure were the Young's modulus, shear modulus, Poisson's ratio and linear compressibility. In Ref. [25], we showed that except for the Poisson's ratio, the dependence on $n$ of the other three elastic properties of GYs could be modelled as a simple serial association of $n$ springs. Here, we extend the analysis of that data investigating how the GY elastic properties depend on the density of the structures. The nomenclature used in Ref. [24] is also adopted for the sake of simplicity, where each GY is named as "G$n$Y$f$". $n$ and $f$ are the number of acetylene linkages in the chains, $1 \leq n \leq 10$, and the corresponding GY family or type, $1 \leq f \leq 7$, respectively, according to Ivanovskii notation [2]. Figure **1** shows the structures of the 7 GY families with $n = 1$.

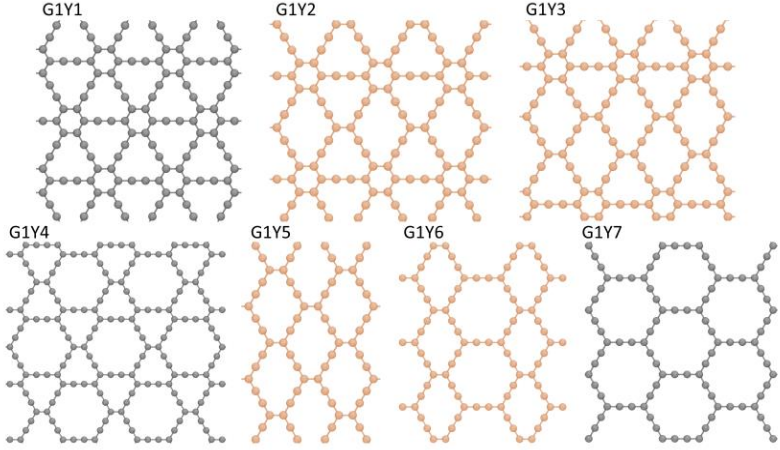

**Figure 1.** The seven GY structures proposed by Baughman, Eckhardt and Kertesz [1]. Horizontal ($x$) and vertical ($y$) axes are related to "armchair" and "zigzag" GY directions, respectively. The numbering notation and sequence are the same as defined by us in Ref. [24]. Light gray (light orange) structures are symmetric (non-symmetric).

In the next section, we will describe the theory and the MD simulations details. Then, the results and discussions are presented followed by the conclusions section.

## THEORY AND SIMULATION DETAILS

GY structures were built with sizes varying from about 35 Å (for $n = 1$) to 208 Å (for $n = 10$). The Young's modulus, $E$, the shear modulus, $G$, the linear compressibility, $\beta$, and the Poisson's ratio, $\upsilon$, of all GYs were obtained from the calculation of their in-plane elastic constants $C_{ij}$. These were obtained using the AIREBO potential [25] as available in LAMMPS [26], according to the protocols described in Ref. [24]. Except for the shear modulus, all the quantities were obtained along $x$ and $y$ directions.

The density dependence of the elastic parameters of the structures will be analysed through scaling power-laws of the type:

$$\frac{M_i}{M_{0i}} \propto \left(\frac{\rho}{\rho_0}\right)^b \quad (1)$$

where $M$ could be $E$, $G$, $\beta$ or $\upsilon$, $i = 1$ or $2$ represents the direction $x$ or $y$, respectively, and the index 0 indicates the property of the most dense member of the GY family, i. e., that of the G1Y$f$.

Cellular solids formed by open (closed) cells are predicted to present $b = 2$ ($b = 3$) [27]. Although most of the materials satisfy the open cell prediction, some particular structures deviate from that as aerogels, for example, that present $b = 3.7$ [28]. Other structures present a more complex structure-dependent value of $b$ ($0.6 < b < 4.2$) as porous silica films [29] and graphene nanomeshes [30].

In the present study, it will be interesting to distinguish the GY structures between symmetric (SG) and non-symmetric (NSG). The G$n$Y1, G$n$Y4 and G$n$Y7 families belong to SG group while the G$n$Y2, G$n$Y3, G$n$Y5 and G$n$Y6 ones belong to the NSG group.

## RESULTS AND DISCUSSION

Figure **2** shows the density dependence of the elastic moduli of the seven GY families of structures. The Young's modulus is shown along two directions.

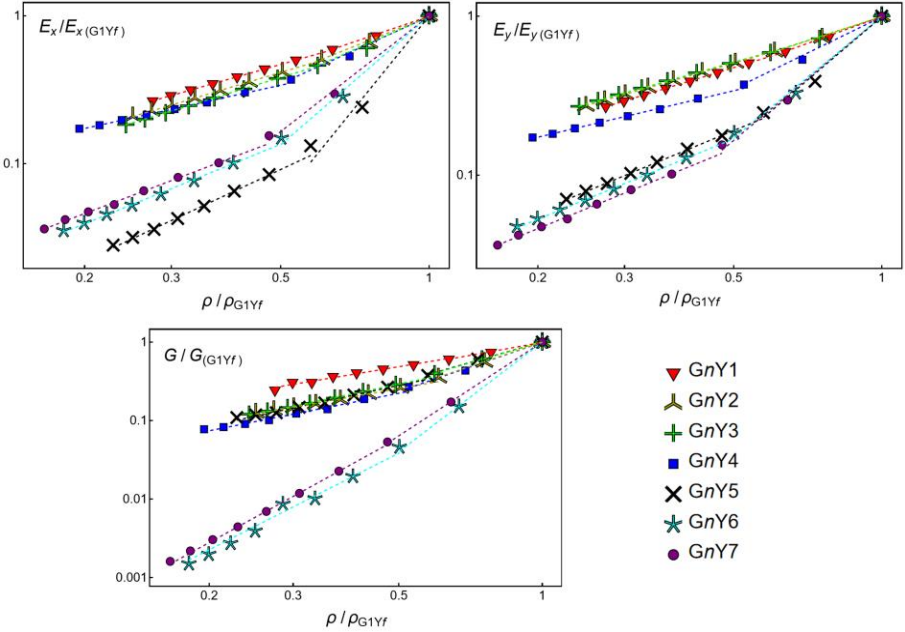

**Figure 2.** Density dependence of the Young's modulus along $x$ and $y$ directions, $E_x$ and $E_y$, respectively, and shear modulus, $G$, of all GYs. For each GY family, a log-log fitting (dashed lines) of Eq. (1) was done at two regions: i) $1 \leq n \leq 3$ and ii) $3 \leq n \leq 10$.

A visual inspection of figure **2** shows that except for G$n$Y1, G$n$Y2 and G$n$Y3 families, there are two clear different trends in the density dependence of the elastic properties of the GY structures. Table **I** shows the fitted values of $b$ for the density dependence of every elastic property of every GY family, except for the Poisson's ratio. The first observation is that for $n \leq 3$, the exponents of the density dependence of the Young's modulus of the GY structures are between ~ 1.0 and 4.1. The exponents are smaller than 2 for the GY families from 1 to 4, and larger than 2 for G$n$Y5, G$n$Y6 and G$n$Y7 families. This indicates that for low $n$, the Young's modulus of the first four GY families are not so sensitive to their density as the last three of them. As mentioned before, exponents smaller (larger) than 2 means that the elastic moduli the structures are less (more) sensitive to the density than cellular materials. In particular, G$n$Y6 and G$n$Y7 families behave like aerogels for low $n$ regarding the density dependence of their Young's moduli. For $n \geq 3$, the exponents of the density dependence of the Young's modulus of the GY structures are between ~ 0.7 and 1.6. It means that the elastic moduli of less dense GYs are much less sensitive to density than that of high dense GYs.

For the shear modulus, its sensitivity to density is also different between low ($n \geq 3$) and high ($n \leq 3$) GY densities. It decreases for large $n$. But differently from the Young's moduli behaviour, the shear moduli of G$n$Y6 and G$n$Y7 families behave like aerogels, being much more sensitive to the structure density than cellular materials, even for large values of $n$. This might be a consequence of the fact that GY structures of G$n$Y6 and G$n$Y7 families have small or zero number of sp2 carbon-carbon bonds, so being less stiff for shearing than the other structures.

Figure 3 shows the density dependence of the linear compressibility and Poisson's ratio of all GY structures.

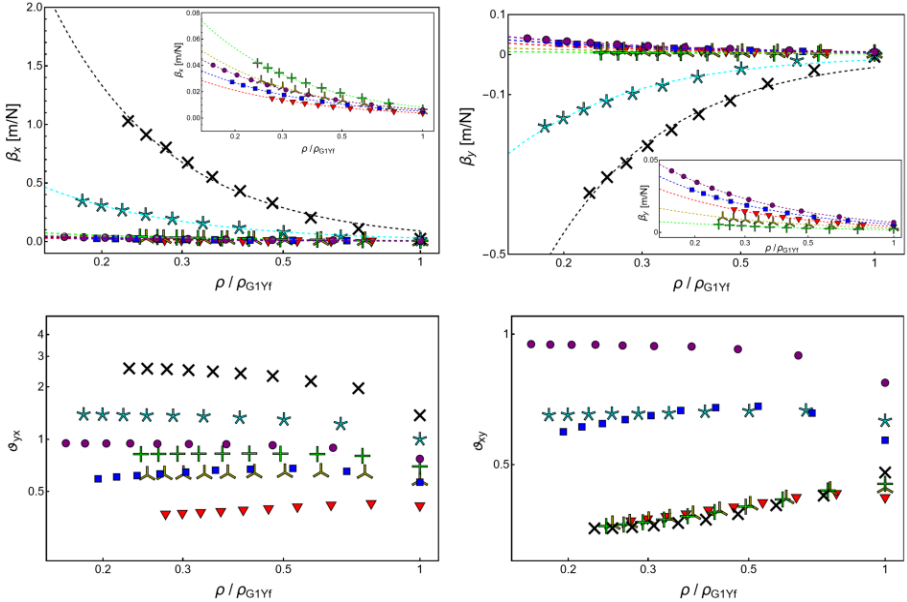

**Figure 3.** Density dependence of the linear compressibility (top) along $x$ and $y$ directions, $\beta_x$ and $\beta_y$, respectively, and Poisson's ratio (bottom) also along $x$ and $y$ directions, $\upsilon_{XY}$ and $\upsilon_{YX}$, respectively, of all GYs. For linear compressibility a fitting curve (dashed lines) of Eq. (1) was done for all values of $n$. Insets show magnifications of the region of low values of linear compressibility. Symbols represent the same GY families as in figure 2.

Except for the structures of the G$n$Y5 and G$n$Y6 families, the linear compressibility of all GYs is roughly proportional to $\sim \rho^{-1}$. In other words, they are inversely proportional to the density. GY structures of G$n$Y5 and G$n$Y6 families present a dependence with $\sim \rho^{-3/2}$. The Poisson's ratio of all GY structures behaves similarly to all materials, i. e., it is roughly constant and does not depend on the density [27].

**Table I.** Values of the exponent $b$ satisfying the Eq. (1) for the density dependence of the Young's modulus, $E$, shear modulus, $G$, and linear compressibility, $\beta$, of every GY family ($1 \leq f \leq 7$) and for two intervals $1 \leq n \leq 3$ and $3 \leq n \leq 10$ for some of them.

|  | G$n$Y1 | G$n$Y2 | G$n$Y3 | G$n$Y4 | G$n$Y5 | G$n$Y6 | G$n$Y7 |
|---|---|---|---|---|---|---|---|
| $E_X$ ($1 \leq n \leq 3$) | 1.14 | 1.33 | 1.46 | 1.51 | 4.11 | 2.89 | 2.60 |
| $E_X$ ($3 \leq n \leq 10$) | 0.95 | 0.99 | 1.02 | 0.74 | 1.56 | 1.35 | 1.27 |
| $E_Y$ ($1 \leq n \leq 3$) | 1.11 | 1.01 | 0.99 | 1.51 | 2.63 | 2.54 | 2.59 |
| $E_Y$ ($3 \leq n \leq 10$) | 0.95 | 0.89 | 0.87 | 0.73 | 1.28 | 1.28 | 1.27 |
| $G$ ($1 \leq n \leq 3$) | 1.09 | 2.04 | 1.76 | 2.08 | 1.74 | 4.64 | 3.97 |
| $G$ ($3 \leq n \leq 10$) | 1.01 | 1.07 | 1.26 | 1.20 | 1.30 | 3.09 | 3.34 |
| $\beta_X$ | -1.06 | -1.13 | -1.15 | -1.02 | -1.67 | -1.43 | -1.01 |
| $\beta_Y$ | -1.05 | -0.93 | -0.74 | -1.02 | -1.63 | -1.54 | -1.01 |

The diversity of values of the exponent of the dependence of the elastic properties of different GYs on density might be related to the geometry of the pores within the structure. Fan *et al*. [29] have shown that films of porous silica having two-dimensional hexagonal pores present $E \sim \rho^{1.0}$. Table **I** shows that most of the GY structures present $0.7 < b < 1.3$. Figure **1** shows that most of the GY structures present hexagonal pores. So, our results are consistent to those of Fan *et al*. The smaller values of *b* like 0.74 for the structures of the G*n*Y4 might be coherent with another result from Fan *et al*. [29]. They have shown that if the geometry of pores in the porous silica film was cubic, the dependence of the elastic modulus with density becomes $E \sim \rho^{0.6}$. As the structures of G*n*Y4 possesses triangular pores in roughly the same proportion to the hexagonal pores, maybe the exponent of its density-dependent elastic modulus became a value in between 0.6 and 1.0. Carpenter *et al*. [30] also observed a large anisotropy in the exponents of the density dependence of the elastic moduli of graphene nanomeshes with holes of different shapes and symmetries.

Our results for the values of the *b* exponents of Eq. (1) are different from what Hernandez and Fonseca [8] have concluded before. They have obtained values of *b* larger than 2. As they have used only GYs with $n = 1$ and $n = 2$, they have no enough amount of data to provide a better result. Besides, they have used the data for all kinds of GY structures, and as we discussed above, the shape of the structure or their structural holes might play an important role on the density-dependence of their elastic moduli.

## CONCLUSION

In this work, unpublished MD results for the density dependence of four elastic properties of ten members of seven GY families are presented. As expected, GY structures from SG (NSG) present symmetric (asymmetric) density dependent elastic properties. We showed that the elastic moduli of the most of the GY structures are less sensitive to density than cellular solids. We also showed that the shear moduli of the structures do not present the same behaviour as function of the density as the Young's moduli. It was attributed to the presence or absence of sp2 carbon-carbon bonds. The density dependence of the linear compressibility of the GYs was presented for the first time. Most of the G*n*Y*f* families were shown to present linear compressibility inversely proportional to the density. The Poisson's ratio of the GY structures behaves as the most of the cellular solids do: it was shown to not depend on the density or, at least, to be weakly dependent on that.

## ACKNOWLEDGMENTS

G. B. K. was supported by the Brazilian agency CNPq. AFF is a fellow of the Brazilian Agency CNPq-Brazil (303284/2021-8) and acknowledges grants #2020/02044-9 from São Paulo Research Foundation (FAPESP) and #2543/22 from FAEPEX/UNICAMP. This work used resources of the John David Rogers Computing Center (CCJDR) in the "Gleb Wataghin" Institute of Physics, University of Campinas.

## CONFLICT OF INTEREST STATEMENT

On behalf of all authors, the corresponding author states that there is no conflict of interest.

## DATA AVAILABILITY STATEMENT

Data available on request from the authors.

## REFERENCES

[1] R. H. Baughman, H. Eckhardt and M. Kertesz, *J. Chem. Phys.* **87**, 6687 (1987). https://doi.org/10.1063/1.453405
[2] A. L. Ivanovskii, *Progress in Solid State Chemistry* **41**, 1 (2013). https://doi.org/10.1016/j.progsolidstchem.2012.12.001
[3] Y. Li, L. Xu, H. Liu and Y. Li, *Chem. Soc. Rev*. **43**, 2572 (2014). https://doi.org/10.1039/c3cs60388a
[4] S. W. Cranford and M. J. Buehler, *Carbon* **49**, 4111 (2011). https://doi.org/10.1016/j.carbon.2011.05.024
[5] J. E. Padilha, A. Fazzio and A.J.R. da Silva, *J. Chem. Phys. C* **118**, 18793 (2014). https://doi.org/10.1021/jp5062804
[6] D. Galhofo and N. Silvestre, *Mechanics of Advanced Materials and Structures* **28**, 495 (2021). https://doi.org/10.1080/15376494.2019.1578007
[7] X. –M. Wang, D. –C. Mo and S. –S. Lu, *J. Chem. Phys*. **138**, 204704 (2013). https://doi.org/10.1063/1.4806069
[8] S. A. Hernandez and A. F. Fonseca, *Diamond and Related Materials* **77**, 57 (2017). https://doi.org/10.1016/j.diamond.2017.06.002
[9] N. Narita, S. Nagai, S. Suzuki and K. Nakao, *Phys. Rev. B* **58**, 11009 (1998). https://doi.org/10.1103/PhysRevB.58.11009
[10] G. Li, Y. Li, H. Liu, Y. Guo, Y. Li and D. Zhu, *Chem. Commun.* **46**, 3256 (2010). https://doi.org/10.1039/b922733d
[11] R. Matsuoka, R. Sakamoto, K. Hoshiko, S. Sasaki, H. Masunaga, K. Nagashio and H. Nishihara, *J. Am. Chem. Soc.* **139**,3145 (2017). https://doi.org/10.1021/jacs.6b12776
[12] K. Khan, A. K. Tareen, M. Iqbal, Z. Shi, H. Zhang and Z. Guo, *Nano Today* **39**, 101207 (2021). https://doi.org/10.1016/j.nantod.2021.101207
[13] J. Gao, J. Li, Y. Chen, Z. Zuo, Y. Li, H. Liu and Y. Lia, *Nano Energy* **43**, 192 (2018). https://doi.org/10.1016/j.nanoen.2017.11.005
[14] Q. Pan, S. Chen, C. Wu, F. Shao, J. Sun, L. Sun, Z. Zhang, Y. Man, Z. Li, L. He and Y. Zhao, *CCS Chem* **2**, 1368 (2020). https://doi.org/10.31635/ccschem.020.202000377
[15] Y. Hu, C. Wu, Q. Pan, Y. Jin, R. Lyu, V. Martinez, S. Huang, J. Wu, L. J. Wayment, N. A. Clark, M. B. Raschke, Y. Zhao and W. Zhang, *Nat. Synth*. **1**, 449 (2022). https://doi.org/10.1038/s44160-022-00068-7
[16] X. Liu, S. M. Cho, S. Lin, Z. Chen, W. Choi, Y. -M. Kim, E. Yun, E. H. Baek, D. H. Ryu and H. Lee, *Matter* **5**, 2306 (2022). https://doi.org/10.1016/j.matt.2022.04.033
[17] V. G. Desyatkin, W. B. Martin, A. E. Aliev, N. E. Chapman, A. F. Fonseca, D. S. Galvão, E. R. Miller, K. H. Stone, Z. Wang, D. Zakhidov, F. T. Limpoco, S. R. Almahdali, S. M. Parker, R. H. Baughman and V. O. Rodionov, *J. Am. Chem. Soc.* **144**, 17999 (2022). https://doi.org/10.1021/jacs.2c06583
[18] B. Zhang, S. Wu, X. Hou, G. Li, Y. Ni, Q. Zhang, J. Zhu, Y. Han, P. Wang, Z. Sun and J. Wu, *Chem* **8**, 2831 (2022). https://doi.org/10.1016/j.chempr.2022.08.002
[19] Z. Yang, Y. Zhang, M. Guo and J. Yun, *Comput. Mater. Sci*. **160**, 197 (2019). https://doi.org/10.1016/j.commatsci.2018.12.033
[20] S. Lin and M. J. Buehler, *Nano* **5**, 11801 (2013). https://doi.org/10.1039/C3NR03241H
[21] C. Esmaeili, J. Song, Y. Li, L. Mao, H. Liu, F. Wu, Y. Chen, C. Liu and X. -E. Zhang, *J. Electrochem. Soc*. **168**, 077520 (2021). https://doi.org/10.1149/1945-7111/ac139c
[22] N. Liang, X. Wu, Y. Lv, J. Guo, X. Zhang, Y. Zhu, H. Liu and D. Jia, *J. Mater. Chem. C* **10**, 2821 (2022). https://doi.org/10.1039/D1TC04406K
[23] Z. Zhu, Q. Bai, S. Li, S. Li, M. Liu, F. Du, N. Sui, W. W. Yu, *Small* **16**, 2001440 (2020). https://doi.org/10.1002/smll.202001440
[24] G. B. Kanegae and A. F. Fonseca, *Carbon Trends* **7**,100152 (2022). https://doi.org/10.1016/j.cartre.2022.100152

[25] D. W. Brenner, O. A. Shenderova, J. A. Harrison, S. J. Stuart, B. Ni and S. B. Sinnott, *J. Phys.: Condens. Matter* **14**, 783 (2002). https://doi.org/10.1088/0953-8984/14/4/312
[26] A. P. Thompson, H. M. Aktulga, R. Berger, D. S. Bolintineanu, W. M Brown, P. S. Crozier, P. J. in 't Veld, A. Kohlmeyer, S. G. Moore, T. D. Nguyen, R. Shan, M. J. Stevens, J. Tranchida, C. Trott, and S. J. Plimpton, *Comput. Phys. Commun.* **271**, 108171 (2022). https://doi.org/10.1016/j.cpc.2021.108171
[27] I. J. Gibson and M. F. Ashby, *Proc. R. Soc. Lond. A* **382**, 43 (1982). http://doi.org/10.1098/rspa.1982.0088
[28] T. Woignier and J. Phalippou, *J. Phys. Colloques* **50**, C4-179 (1989). https://doi.org/10.1051/jphyscol:1989429
[29] H. Fan, C. Hartshorn, T. Buchheit, D. Tallant, R. Assink, R. Simpson, D. J. Kissel, D. J. Lacks, S. Torquato and C. J. Brinker, *Nat. Mater.* **6**, 418 (2007). https://doi.org/10.1038/nmat1913
[30] C. Carpenter, A. M. Christmann, L. Hu, I. Fampiou, A. R. Muniz, A. Ramasubramaniam and D. Maroudas, *Appl. Phys. Lett*. **104**, 141911 (2014); https://doi.org/10.1063/1.4871304